\documentclass[preprint,showpacs,preprintnumbers,amsmath,amssymb]{revtex4}

\usepackage{graphicx}% Include figure files
\usepackage{dcolumn}% Align table columns on decimal point
\usepackage{bm}% bold math
\usepackage{latexsym}
\usepackage{amsfonts}
\usepackage{amssymb}
\usepackage{amsmath}
\usepackage[usenames]{color}
\begin{document}
\preprint{KUNS 2288}

\title{Catastrophic Instability of Small Lovelock Black Holes }

\author{Tomohiro Takahashi}
\author{Jiro Soda}
\affiliation{Department of Physics,  Kyoto University, Kyoto, 606-8502, Japan
}

\date{\today}% It is always \today, today,
             %  but any date may be explicitly specified

%===============================================================%
%************************* ABSTRACT ****************************%
%===============================================================%
\begin{abstract}
We study the stability of static black holes in Lovelock theory which is a natural higher dimensional generalization of Einstein theory. 
We show that Lovelock black holes are stable under vector perturbations
in all dimensions. However, we prove that small Lovelock black holes are unstable
 under tensor perturbations in even-dimensions and 
under scalar perturbations in odd-dimensions.
Therefore, we can conclude that small Lovelock black holes are unstable in any dimensions.
The instability is stronger on small scales and hence catastrophic
in the sense that there is no smooth descendant.  
\end{abstract}

% PACS, the Physics and Astronomy
                             % Classification Scheme.
%\keywords{Suggested keywords}%Use showkeys class option if keyword
                              %display desired
\maketitle

%===============================================================%
%************************ SECTION I ****************************%
%===============================================================%
\section{Introduction}

The possibility of higher dimensional black hole creation at the LHC
would be the most fascinating prediction of
 the braneworld with large extra-dimensions~\cite{Giddings:2001bu}.
 Nowadays, it is fashionable to study higher dimensional black holes.
The study of higher dimensional black holes is not a straightforward
extension of the 4-dimensional one. In fact, in higher dimensions, there are
many black objects with non-trivial topologies such as black rings. 
This seems to require the stability analysis of higher dimensional black objects. 
Moreover, in higher dimensions, Einstein theory is not the most general theory
of gravity which contains terms only up to the second order derivatives
in the equations of motion. 
The most general theory in this sense is Lovelock theory~\cite{Lovelock:1971yv}.
Thus, it is worth extending the stability analysis to
this general Lovelock theory. 

The stability of higher dimensional black holes has been 
intensively studied since the seminal papers by Kodama and Ishibashi~\cite{Kodama:2003jz}.
It is important to study various black holes in Einstein theory
because black holes produced at the LHC
are expected to be charged or rotating. 
A numerical study of charged black holes has been done~\cite{Konoplya:2007jv}.
To investigate the stability of rotating black holes,
a group theoretical method is developed~\cite{Murata:2007gv}.
The method is used to study the stability of squashed black
 holes~\cite{Kimura:2007cr,Ishihara:2008re,Nishikawa:2010zg}
 and 5-dimensional rotating black holes with equal angular momenta~\cite{Murata:2008yx}.
The stability of a special class of rotating black holes in more than 5-dimensions is also
studied~\cite{Kunduri:2006qa,Oota:2008uj,Kodama:2008rq}.
Recent development of numerical stability analysis is 
remarkable~\cite{Dias:2009iu,Yoshino:2009xp,Dias:2010eu,
Shibata:2010wz,Dias:2010maa}.
It has been shown that rapidly rotating black holes are unstable. 

As we mentioned already, in higher dimensions, we need to extend Einstein
theory to Lovelock theory. In addition to the theoretical requirement,
we have a physical motivation to consider higher derivative theory of gravity.
In fact, at the energy scale of black hole production,
Einstein theory is not reliable any more. 
It is believed that string theory which can be consistently formulated 
only in 10-dimensions is the most promising candidate of the unified theory.
We should recall that string theory predicts Einstein theory
only in the low energy limit~\cite{Boulware:1985wk}. In string theory,
  there are higher curvature corrections in addition to the
Einstein-Hilbert term~\cite{Boulware:1985wk}. Thus, it 
is natural to extend gravitational theory into those
 with higher power of curvature in higher dimensions.
  It is Lovelock theory 
that belongs to such class of theories~\cite{Lovelock:1971yv,Christos}.
In Lovelock theory, it is known that
there exist static spherical symmetric black hole solutions~\cite{Wheeler:1985nh} 
(and topological black hole solutions are also found in~\cite{Cai:2001dz}). 
Hence, it is natural to suppose black holes produced at the LHC are 
of this type~\cite{Barrau:2003tk}.
Thus, it is important to study the stability of these Lovelock black holes.

In the case of the second order Lovelock theory, the so-called 
Einstein-Gauss-Bonnet theory, the stability analysis under tensor
perturbations has been performed~\cite{Dotti:2004sh} (see also an earlier work~\cite{Neupane:2003vz}). 
The analysis has been also extended to the scalar and vector 
perturbations~\cite{Gleiser:2005ra}.
It is shown that there exists the scalar mode instability in 5-dimensions, 
the tensor mode instability in 6-dimensions, 
and no instability in other dimensions.
Although Einstein-Gauss-Bonnet theory is the most general theory
in 5 and 6-dimensions, it is not so in more than 6-dimensions.
For example, when we consider 10-dimensional black holes, 
we need to incorporate the fourth order Lovelock term. 
Indeed, when we consider black holes at the LHC, 
it is important to consider these higher order Lovelock terms~\cite{Rychkov:2004sf}.
Hence, the purpose of this paper is to study the stability of black holes in any order
Lovelock theory, namely, in any dimensions. 
We have already shown that 
Lovelock black holes are unstable in even-dimensions under tensor perturbations~\cite{Takahashi:2009dz}. 
In this paper, we extend previous results to vector and scalar perturbations
using the master equations we have obtained recently~\cite{Takahashi}.

The organization of this paper is as follows.
 In section \ref{seq:2}, we review Lovelock theory and 
explain a graphical method for constructing Lovelock black hole solutions. 
In section \ref{seq:3}, we consider tensor perturbations and show the instability of small Lovelock black holes in even-dimensions. 
This is a review of our previous paper~\cite{Takahashi:2009dz}. 
In section \ref{seq:4}, we show that black holes are stable under vector perturbations.
In section \ref{seq:5}, we examine scalar perturbations and show that 
there exists the instability in odd-dimensions if black holes are sufficiently small.  
In section \ref{seq:6}, we present a detailed analysis for Einstein-Gauss-Bonnet
theory to illustrate our statements.
 The final section \ref{seq:7} is devoted to the conclusion.

%===============================================================%
%************************ SECTION II ****************************%
%===============================================================%
\section{Lovelock Black Holes}
\label{seq:2}

In this section, we review Lovelock theory and introduce
a graphical method to reveal the nature of asymptotically flat
black hole solutions. 

 The most general  divergence free symmetric tensor constructed out of 
 a metric and its first and second derivatives has been obtained
  by Lovelock~\cite{Lovelock:1971yv}.
The corresponding Lagrangian can be constructed from $m$-th order Lovelock terms
\begin{eqnarray}
  {\cal L}_m = \frac{1}{2^m} 
  \delta^{\lambda_1 \sigma_1 \cdots \lambda_m \sigma_m}_{\rho_1 \kappa_1 \cdots \rho_m \kappa_m}
  R_{\lambda_1 \sigma_1}{}^{\rho_1 \kappa_1} \cdots  R_{\lambda_m \sigma_m}{}^{\rho_m \kappa_m}
                       \ ,
\end{eqnarray}
where  $R_{\lambda \sigma}{}^{\rho \kappa}$ is the Riemann tensor in $D$-dimensions
and $\delta^{\lambda_1 \sigma_1 \cdots \lambda_m \sigma_m}_{\rho_1 \kappa_1 \cdots \rho_m \kappa_m}$ is the 
generalized totally antisymmetric Kronecker delta.
By construction, the Lovelock terms vanish for $2m>D$. It is also known that
the Lovelock term with $2m=D$ is a topological term.   
Thus, Lovelock Lagrangian in  $D$-dimensions is defined by
\begin{eqnarray}
  L = \sum_{m=0}^{k} c_m {\cal L}_m \ ,   \label{eq:lag}
\end{eqnarray}
where we defined the maximum order $k\equiv [(D-1)/2]$ and  $c_m$ are 
arbitrary constants. 
Here, $[z]$ represents the maximum integer satisfying $[z]\leq z$. 
Hereafter, we set $c_0=-2\Lambda$, $c_1=1$ and $c_m=a_m/m\ (m\geq 2)$ for convenience. 
Taking variation of the Lagrangian with respect to the metric,
 we can derive Lovelock equation
\begin{eqnarray}
	0={\cal G}_{\mu}^{\nu}
      =\Lambda \delta_{\mu}^{\nu}-\sum_{m=1}^{k}\frac{1}{2^{(m+1)}}\frac{a_m}{m} 
	 \delta^{\nu \lambda_1 \sigma_1 \cdots \lambda_m \sigma_m}_{\mu \rho_1 \kappa_1 \cdots \rho_m \kappa_m}
       R_{\lambda_1 \sigma_1}{}^{\rho_1 \kappa_1} \cdots  R_{\lambda_m \sigma_m}{}^{\rho_m \kappa_m}  \ . \label{eq:EOM}
\end{eqnarray}

As is shown in \cite{Wheeler:1985nh, Cai:2001dz}, there exist static 
exact solutions of Lovelock equation. 
Let us consider the following metric
\begin{eqnarray}
   ds^2=-f(r)dt^2 + \frac{dr^2}{f(r)}+r^2{\bar \gamma}_{i j}dx^idx^j \ ,\label{eq:solution}
\end{eqnarray}
where ${\bar \gamma}_{ij}$ is the metric of $n\equiv D-2$-dimensional constant
 curvature space with a curvature $\kappa$=1,0 or -1.  
Using this metric ansatz, we can calculate Riemann tensor components as
\begin{eqnarray}
	R_{tr}{}^{tr}=-\frac{f^{''}}{2},\ R_{ti}{}^{tj}=R_{ri}{}^{rj}=-\frac{f^{'}}{2r}\delta_{i}{}^{j},\ R_{ij}{}^{kl}=\left(\frac{\kappa-f}{r^2}\right)\left(\delta_{i}{}^{k}\delta_{j}{}^{l}-\delta_{i}{}^{l}\delta_{j}{}^{k}\right) \ . \label{eq:riemann}
\end{eqnarray}
Substituting (\ref{eq:riemann}) into (\ref{eq:EOM}) and defining a new variable
$\psi(r)$ by
\begin{eqnarray}
       f(r)=\kappa-r^2\psi(r) \ , \label{eq:def}
\end{eqnarray}
we obtain an algebraic equation 
\begin{eqnarray}
       W[\psi]\equiv\sum_{m=2}^{k}\left[\frac{a_m}{m}\left\{\prod_{p=1}^{2m-2}(n-p)\right\}\psi^m\right]+\psi-\frac{2\Lambda}{n(n+1)}=\frac{\mu}{r^{n+1}}  \ .
\label{eq:poly}
\end{eqnarray}
In (\ref{eq:poly}), we used $n = D-2$ and $\mu$ is a constant of integration
 which is related to the ADM mass as~\cite{Myers:1988ze}:
\begin{eqnarray}
	M=\frac{2\mu\pi^{(n+1)/2}}{\Gamma((n+1)/2)} \ , \label{eq:ADM}
\end{eqnarray}
where we used a unit $16\pi G=1$.

\begin{figure}[h]
		  \begin{center}
		    \includegraphics[height=7cm, width=11cm]{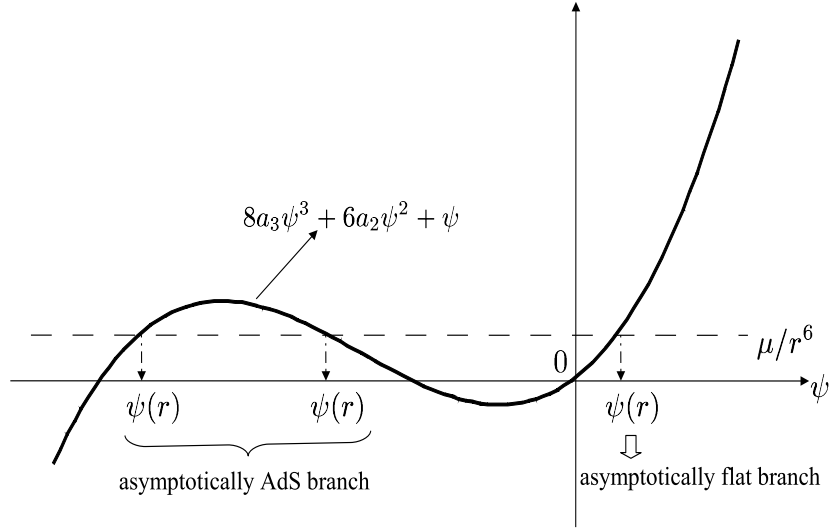}
		    \caption{We illustrate a graphical method for $n=5$ case.
                 In this case, the third order Lovelock theory
		             is most general. Therefore, $W[\psi]$ in (\ref{eq:poly})
		 reads a cubic polynomial. 
             In this figure, three roots are depicted. Among these roots, 
		             only $\psi\geq 0$ one corresponds to an
                          asymptotically flat solution.}
		   \label{fig:n5} 
		  \end{center} 
	\end{figure}

From (\ref{eq:poly}), it is easy to see that the solution $f(r)$ has many branches.
 In this paper, we want to concentrate on asymptotically flat 
spherically symmetric, i.e. $\Lambda=0$ and $\kappa=1$, solutions 
with a positive ADM mass  $\mu>0$
  because such black holes could be created at the LHC. 
We also assume that Lovelock coefficients satisfy
\begin{eqnarray}
	a_m\geq 0 \ , \label{eq:conditions}
\end{eqnarray}
for simplicity. 
	\begin{figure}
		    \begin{center}
		      \includegraphics[height=8cm, width=12cm]{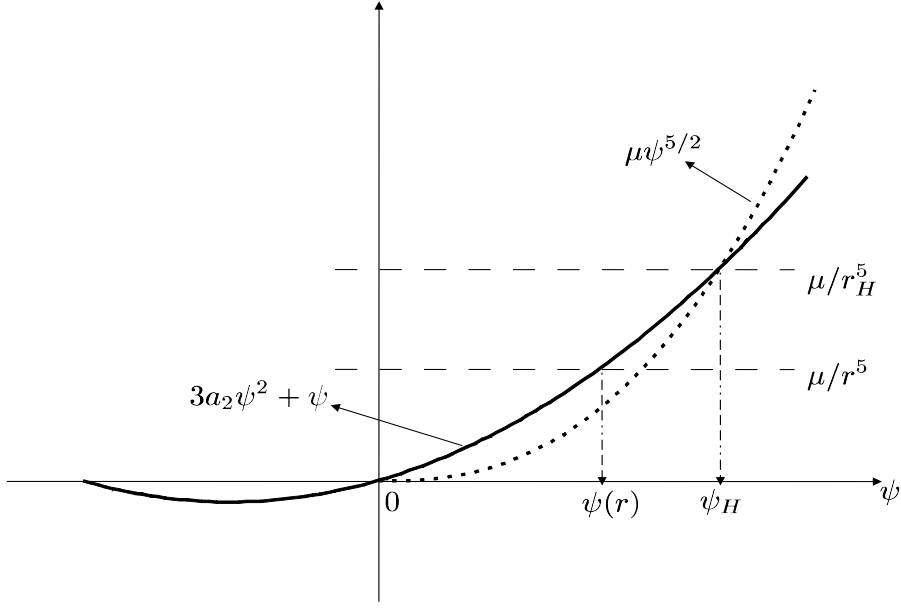}
		      \caption{The intersection between 
                  the solid curve and the thin horizontal line determines
                  the solution $\psi = \psi (r)$ for the case $n=4$.
                  Apparently, the infinity $r=\infty$ corresponds to $\psi =0$.
                  The intersection between solid and dashed curve
                  gives a horizon $r_H$.  }
		       \label{fig:1}
		     \end{center}
      \end{figure}
For example, 
in the case of $n=3$, the theory is reduced to Einstein-Gauss-Bonnet
theory. In this case,  Eq.(\ref{eq:poly}) reads
\begin{eqnarray}
  a_2 \psi^2 + \psi = \frac{\mu}{r^4} \ ,
\end{eqnarray}
which can be explicitly solved as
\begin{eqnarray}
\psi = \frac{-1\pm \sqrt{1+\frac{4 a_2 \mu}{r^4}}}{2a_2} \ .
\end{eqnarray}
The upper branch leads 
to the asymptotically flat solution
\begin{eqnarray}
  f(r) = 1 + \frac{r^2}{2a_2} 
  \left[ 1 - \sqrt{1+\frac{4a_2\mu}{r^4}}\right]  \ .
\end{eqnarray}
In the case of $n=4$, Eq.(\ref{eq:poly}) is reduced to
\begin{eqnarray}
  3 a_2 \psi^2 +\psi = \frac{\mu}{r^5} \ .
\end{eqnarray}
Although it is easy to solve this equation analytically,
in Fig.\ref{fig:1}, a graphical method is also explained for this case.
However, in the case of $n=5$, Eq.(\ref{eq:poly}) 
becomes the third order algebraic equation
\begin{eqnarray}
8 a_3 \psi^3 +6a_2 \psi^2 +\psi = \frac{\mu}{r^6} \ .
\label{eq:poly5}
\end{eqnarray}
We have a formula for solutions of Eq.(\ref{eq:poly5}),
however, the roots are complicated in general. 
Hence, we use a graphical method illustrated in Fig.\ref{fig:n5}. 
Because of the conditions (\ref{eq:conditions}),
the function is monotonic for positive $\psi$. 
From (\ref{eq:poly}), we see the root behaves $\psi \sim \mu/r^{n+1}$ 
or $f(r)\sim 1-\mu/r^{n-1}$ as $r\rightarrow \infty$.
Thus, the asymptotically flat solutions belong to the branch where $\psi$ 
is always positive. 
 
\begin{figure}[t]
		  \begin{center}
		    \includegraphics[height=8cm, width=12cm]{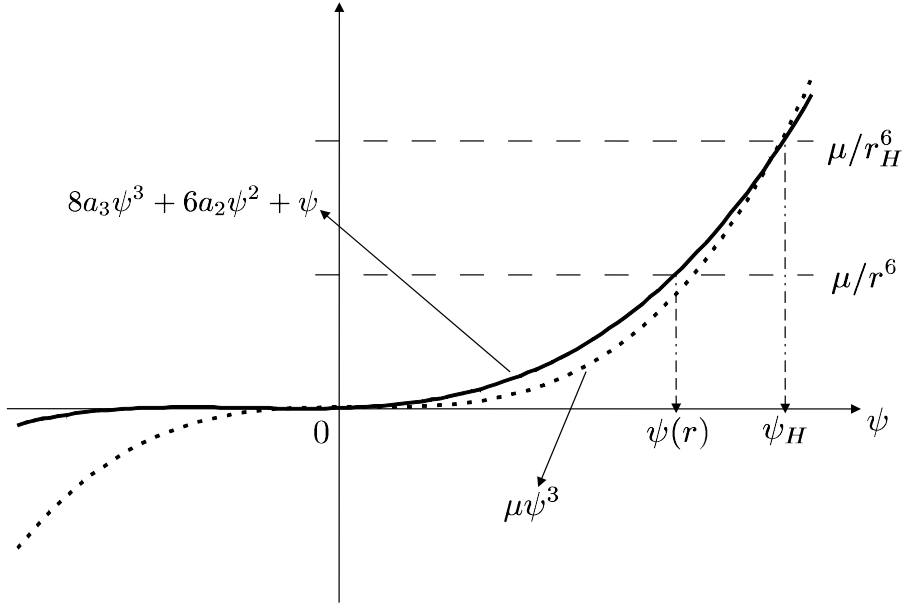}
		    \caption{For $n=5$ case, this figure explains a
                method for calculating $\psi_H$ or $r_H$ graphically.
                The positive root of Eq.(\ref{eq:rh}) $\psi_H$ can be obtained from 
                the intersection of the solid and dashed curves.
                Since the intersection between the horizontal line and the solid curve
                gives a solution $\psi =\psi (r)$, the horizon radius
                $r_H$ is determined from the
                intersection of the horizontal line, the solid and dashed curves.}
		   \label{fig:2} 
		  \end{center} 
	\end{figure}

Now, let us look for the horizon of black holes.
The horizon 
radius of the asymptotically flat solution is characterized  by $f(r_H)=0$.
 From (\ref{eq:def}), we have a relation $\psi_H\equiv \psi(r_H)=1/r_H^2$.
  Using this relation and (\ref{eq:poly}), we obtain an algebraic equation
\begin{eqnarray}
              W[\psi_H]  =\mu \psi_H^{(n+1)/2} \ .
 \label{eq:rh}
\end{eqnarray}	
This determines $\psi_H$ and hence $r_H$. 
For example, in the case of $n=3$, we can easily solve this polynomial equation
(\ref{eq:rh}) as 
\begin{eqnarray}
\psi_H=\frac{1}{\mu-a_2}\ .
\label{eq:rh1}
\end{eqnarray}
Note that if $\mu \leq a_2$, there appears a naked singularity. 
So, we consider the range $\mu>a_2$.  
In the case of n=4, since 
it is a bit complicated to solve Eq.(\ref{eq:rh}) analytically, 
we present a graphically method in Fig.\ref{fig:1}. 
Similarly, in Fig.\ref{fig:2}, 
we present a graphical method to solve Eq.(\ref{eq:rh}) for $n=5$. 
 From Fig.\ref{fig:2}, it is obvious  that the range 
 $\infty\geq r \geq r_H$ corresponds to
$0\leq \psi \leq \psi_H$ when $f(r)$ describes an asymptotically flat solution. 
It is also apparent that $\psi_H$ becomes larger as $\mu$ becomes smaller.  

Remarkably, the nature of black holes depends on the dimensions.
In even-dimensions $n=2k$, dividing (\ref{eq:rh}) by $\psi_H$, we have 
\begin{eqnarray}
-\mu\psi_H^{(2k-1)/2}+\sum_{m=2}^{k}\left[\frac{a_m}{m}\left\{\prod_{p=1}^{2m-2}(n-p)\right\}\psi_H^{m-1}\right]+1=0\ .
\label{}
\end{eqnarray}
Near $\psi_H=0$, the left hand side is positive; however, when $\psi_H$ is sufficiently large,
 it is negative. Then, there exists a positive root somewhere between. 
This root moves from $0$ to $\infty$ as $\mu$ moves from $\infty$ to $0$.
Thus, there is no restriction for $\mu$ in this case. 
On the other hand, in odd-dimensions $n=2k-1$, 
after dividing by $\psi_H$, Eq.(\ref{eq:rh}) becomes 
\begin{eqnarray}
\left(\frac{a_k}{k}\left\{\prod_{p=1}^{2k-2}(n-p)\right\}-\mu\right)\psi_H^{k-1}+\sum_{m=2}^{k-1}\left[\frac{a_m}{m}\left\{\prod_{p=1}^{2m-2}(n-p)\right\}\psi_H^{m-1}\right]+1=0\ .
\label{psi_H_cal}
\end{eqnarray}
In order to have a positive root, we need 
 $\mu>\frac{a_k}{k}\left\{\prod_{p=1}^{2k-2}(n-p)\right\}$. 
 Hence, we have the lower bound for the mass in odd-dimensions.
In fact, there exists a positive root in this case,
because the left hand side of Eq.(\ref{psi_H_cal}) is positive near $\psi_H=0$ and 
negative for sufficiently large $\psi_H$. Furthermore, It is not difficult to see
 that this root approaches $0$ as $\mu\rightarrow \infty$ and approaches $\infty$ as 
 $\mu\rightarrow\frac{a_k}{k}\left\{\prod_{p=1}^{2k-2}(n-p)\right\}$. 
Therefore, the root $\psi_H$ moves in the range $0<\psi_H<\infty$ as $\mu$ moves
in the range 
$\frac{a_k}{k}\left\{\prod_{p=1}^{2k-2}(n-p)\right\}< \mu<\infty$. 

Finally, we examine the singularity in the solutions. 
Using the metric ansatz (\ref{eq:solution}), the Kretschmann scalar $R_{\mu \nu \lambda \rho}R^{\mu \nu \lambda \rho}$ is 
\begin{eqnarray}
	R_{\mu \nu \lambda \rho}R^{\mu \nu \lambda \rho}=f^{''}+2n\frac{f^{'2}}{r^2}+2n(n-1)\frac{(\kappa-f)^2}{r^4}\ .\nonumber
\end{eqnarray}
Thus, this solution has curvature singularities at $r=0$
 or at the point where derivatives of $f(r)$ diverges.  
For example, $f^{'}$ diverges at the point where $\psi^{'}$ diverges 
because of $f^{'}=-2r\psi+r^2\psi^{'}$. 
Taking a derivative of (\ref{eq:poly}) with respect to $r$, we have
a relation 
\begin{eqnarray}
\psi^{'}=-\frac{(n+1)\mu}{r^{n+2}\partial_{\psi}W[\psi]}\ .\nonumber
\label{}
\end{eqnarray}
Hence, a curvature singularity appears at the point $\partial_{\psi}W=0$. 
However, for the asymptotically flat branch, 
$W[\psi]$ is monotonically increasing function of $\psi$.
 Hence, we conclude $\partial_{\psi}W[\psi]>0$.   
Similarly, for asymptotically flat spherical symmetric solutions, 
we can prove that the conditions $|1-f|<\infty$, $|f^{'}|<\infty$ and $|f^{''}|<\infty$
 are satisfied except for $r=0$. Therefore, 
there is a curvature singularity only at $r=0$ in the cases we are considering.
 And asymptotically flat solutions have the horizon 
if the parameter $\mu$ is sufficiently large. 
So, these solutions do not have a naked singularity and 
describe black holes with a mass $M$ defined by Eq.(\ref{eq:ADM}).  

%===============================================================%
%************************ SECTION III ****************************%
%===============================================================%

\section{stability analysis for tensor perturbations}
\label{seq:3}

In this section, we examine the stability under tensor  perturbations, 
which we have already studied in \cite{Takahashi:2009dz}. 

We start from the master equation for tensor perturbations
\begin{eqnarray}
	- f^2\chi^{''}
      - \left( f^2 \frac{T^{''}}{T^{'}}+ \frac{2f^2}{r}
      +  f f^{'} \right) \chi^{'}
      +  \frac{(2\kappa+\gamma_t)f}{(n-2)r}\frac{T^{''}}{T^{'}} \chi 
      =   \omega^2 \chi \ ,
 \label{tensor_master}
\end{eqnarray}
where $\omega$ is a frequency and we have defined a key function
\begin{eqnarray}
T(r)=r^{n-1}\partial_{\psi}W[\psi]\ .
\label{}
\end{eqnarray}
In (\ref{tensor_master}), $\chi $ is the master variable 
and $\gamma_t$ is eigenvalue of tensor harmonics which is given by 
$\gamma_t =\ell (\ell +n-1)-2$, ($\ell =2,3,4 \cdots$) for $\kappa=1$ 
and  positive real numbers for $\kappa=-1,0$.

 Here, we should recall our assumptions.
  We assumed the conditions (\ref{eq:conditions}) are satisfied. 
  And we also assumed spherical symmetry and positivity of the mass,
   i.e., $\kappa=1$ and $\mu>0$.
  Then, there exists an asymptotically flat spherical symmetric branch which we have considered in section 
  \ref{seq:2}. Note that $T(r)$ which is defined above is always positive in this branch. 
   
\subsection{Criterion for Stability}

 We will present the condition for the stability of the solutions we are considering. 
   
As we will soon see, the master equation (\ref{tensor_master}) can be transformed
 into the Schr${\rm {\ddot o}}$dinger form. 
To do this, we have to impose the condition 
\begin{eqnarray}
	T^{'}(r)>0 \ ,  \quad ({\rm for}\ r>r_H) \ . \label{eq:assum}
\end{eqnarray}
In fact, this is necessary for the linear analysis to be applicable. 
 In the case that there exists $r_0$ such that $T^{'}(r_0)=0$ and $r_0>r_H$,
 we encounter a singularity. 
 Using approximations $T^{'}(r)\sim T^{''} (r_0)(r-r_0)\equiv T^{''}(r_0) y$, 
$f(r)= f(r_0 ) $ and $r=r_0$, (\ref{tensor_master}) approximately becomes 
\begin{eqnarray}
	y \frac{d^2\chi}{dy^2}+\frac{d\chi}{dy}+c \chi=0  \ .
\end{eqnarray}
This  shows that near $r=r_0$, $\chi$ behaves as  $\chi \sim c_1 + c_2 \log y$ , 
where $c_1$ and $c_2$ are constants of integration. 
Hence, the solution is singular at $y =0$ for generic perturbations. 
 The similar situation occurs even in cosmology with higher derivative
  terms~\cite{Kawai:1998bn,Satoh:2007gn}. 
 In those cases, this kind of singularity alludes to ghosts.
 Indeed, if there is a region $T^{'}(r)<0$ outside the horizon, the kinetic
 term of perturbations has a wrong sign.
 Hereafter, we call this the ghost instability. 

When the condition (\ref{eq:assum}) is fulfilled, introducing a new variable
$
	\Psi(r)=\chi(r)r\sqrt{T^{'}(r)} 
$
 and switching to the coordinate $r^*$, defined by $dr^*/dr=1/f$,
we can rewrite Eq.(\ref{tensor_master}) as
\begin{eqnarray}
	-\frac{d^2\Psi}{dr^{*2}}+V_t(r(r^*))\Psi=\omega^2\Psi \ , 
      \label{eq:shradinger}
\end{eqnarray}
where
\begin{eqnarray}
	V_t(r)=\frac{(2\kappa+\gamma_t)f}{(n-2)r}\frac{d \ln{T^{'}}}{dr}+\frac{1}{r\sqrt{T^{'}}}f\frac{d}{dr}\left(f\frac{d}{dr}r\sqrt{T^{'}}\right) \label{eq:potential}
\end{eqnarray}
is an effective potential.

For discussing the stability, the "S-deformation" approach
 is useful~\cite{Kodama:2003jz, Dotti:2004sh}.  Let us define the operator 
\begin{eqnarray}
	{\cal H}\equiv -\frac{d^2}{dr^{*2}}+V_t
\end{eqnarray}
acting on smooth functions defined on $I=(r^{*}_H,\infty)$.
Then, (\ref{eq:shradinger}) is the eigenequation and $\omega^2$ is eigenvalue 
of ${\cal H}$. We also define the inner products as
\begin{eqnarray}
	(\varphi_1,\varphi_2)=\int_I \varphi_1^*\varphi_2 dr^* \ .\label{inner_product}
\end{eqnarray}
In this case, for any $\varphi$, we can find a smooth function $S$ such that 
\begin{eqnarray}
	(\varphi,{\cal H}\varphi)=\int_{I} (|D\varphi|^2+\tilde{V}|\varphi|^2)dr^{*},
\end{eqnarray}
where we have defined
\begin{eqnarray}
	D=\frac{d}{dr^{*}}+S  \ , \quad 
	\tilde{V}=V_t+f\frac{dS}{dr}-S^2  \ .\label{tildeV}
\end{eqnarray}
Following~\cite{Dotti:2004sh}, we choose $S$ to be
\begin{eqnarray}
	S=-f\frac{d}{dr}\ln{(r\sqrt{T^{'}})} \ .
\end{eqnarray}
Then, we obtain the formula
\begin{eqnarray}
	(\varphi,{\cal H}\varphi)
      =\int_{I} |D\varphi|^2dr^{*}+(2\kappa+\gamma_t)
      \int_{r_H}^{\infty}\frac{|\varphi|^2}{(n-2)r}\frac{d \ln{T^{'}}}{dr}dr \ .
       \label{eq:stab}
\end{eqnarray}
Here, the point is that the second term in (\ref{eq:stab}) includes
a factor $2\kappa+\gamma_t >0$, but $T^{'}$ does not include $\gamma_t$. 
Hence, by taking a sufficiently large $2\kappa +\gamma_t$, we can 
always make the second term dominant.  

Now, let us show that the sign of $d\ln T^{'} /dr$ determines the stability.
If $d\ln T^{'} /dr >0$ on $I$, the solution (\ref{eq:solution}) is stable.
This can be understood as follows. 
Note that $2\kappa+\gamma_t>0$, 
then we have $\tilde{V}>0$ for this case.
 That means $(\varphi,{\cal H}\varphi)>0$ for arbitrary $\varphi$ if $d \ln{T^{'}}/dr>0$ on $I$. 
We choose, for example, $\varphi$ as the lowest eigenstate, then we can conclude that the lowest eigenvalue $\omega^2_0$ is positive. 
Thus, we proved the stability. The other way around,
if $d \ln{T^{'}} /dr <0$ at some point in $I$, the solution is unstable. 
To prove this, the inequality
\begin{eqnarray}
	\frac{(\varphi,{\cal H}\varphi)}{(\varphi,\varphi)} \geq \omega_0^2
      \label{eq:ineq}
\end{eqnarray}
is useful. This inequality is correct for arbitrary $\varphi$.
 If $d \ln{T'}/dr <0$ at some point in $I$, we can find $\varphi$ such that
\begin{eqnarray}
	 \int_{r_H}^{\infty}\frac{|\varphi|^2}{(n-2)r}\frac{d \ln{T^{'}}}{dr}dr<0 \ .
\end{eqnarray}
In this case, (\ref{eq:stab}) is negative for sufficiently large $2\kappa+\gamma_t$.
 Then, the inequality (\ref{eq:ineq}) implies $\omega^2_0<0$ 
 and the solution has unstable modes. 
Thus, we can conclude that the solution is stable if and only if
 $d \ln{T^{'}}/dr>0$ on $I$.
 
 From the above logic, if $d\ln{T^{'}}/dr$ has a negative region, 
 negative $\omega^2$ states exist. 
 Therefore, this instability is dynamical. Then, we call this as dynamical instability in order to distinguish this from the 
 ghost instability which is caused by negativity of $T^{'}(r)$. 
 
We want to summarize this subsection. 
If $T^{'}$ has negative region outside the horizon $r>r_H$, Lovelock black holes 
have the ghost instability. Even if $T^{'}$ is always positive, 
Lovelock black holes have the dynamical instability 
if $T^{''}$ has a negative region outside the horizon. 
Therefore, Lovelock black holes are stable under tensor perturbations 
if and only if $T^{'}$ and $T^{''}$ are always positive outside the horizon. 

\subsection{Instability of Small Lovelock Black Holes in Even-dimensions}

In this subsection, we check the sign of $T^{'}$ and $T^{''}$ 
for asymptotic flat solutions $f(r)$. 
In order to see the sign of these functions, it is useful to express them  
as functions of $\psi$ instead of $r$. Using Eq.(\ref{eq:poly}) and its derivative,
we obtain 
\begin{eqnarray}
	\left(\partial_{\psi} W[\psi]\right) \psi^{'}
      =-(n+1)\frac{\mu}{r^{n+2}}=-(n+1)\frac{W[\psi]}{r} \ .
\end{eqnarray}
The above formula can be used to eliminate $\psi^{'}$ in $T^{'}$.
The result reads
\begin{eqnarray}
	T^{'}(r)=\frac{r^{n-2}}{\partial_{\psi}W}
   \left[ (n-1)\left(\partial_{\psi}W\right)^2 -(n+1) W\partial_{\psi}^2W \right] 
   \ .\label{T_prime}
\end{eqnarray}
Similarly, $T^{''}$ can be written as
\begin{eqnarray}
	T^{''}&=& \frac{r^{n-3}}{(\partial_\psi W)^3}
      \Biggl[(n-1)(n-2)\left(\partial_{\psi}W\right)^4
      -(n+1)(n-4) W \left(\partial_{\psi}W\right)^2 \partial_{\psi}^2W \nonumber\\
	&\ &\hspace{2.5cm}+(n+1)^2
      W^2\left\{\partial_\psi W \partial_\psi^3 W-(\partial_\psi^2 W)^2\right\}
      \Biggr]\ .\label{T_double_prime} 
\end{eqnarray}
Since $W[\psi]$ is a polynomial function of $\psi$ and $\partial_\psi W$
is positive, we can determine the sign of $T^{'}$ and $T^{''}$ by examining
the sign of polynomial in the numerators. 
Thus, the stability problem has been reduced to an algebraic one.

Substituting the general form of $W[\psi]$ into $T^{'}$, 
we obtain
\begin{eqnarray}
	T^{'}(r)=r^{n-2} \frac{K[\psi]}{1+\sum_{m=2}^{k}\left[a_m\left\{\prod_{p=1}^{2m-2}(n-p)\right\}\psi^{m}\right]}\ , \label{eq:hpsi}
\end{eqnarray}
where
\begin{eqnarray}
	K[\psi]&=&(n-1)+\sum_{m=2}^{k}\left[a_m(n-1)\left\{(3-m)n-(m+1)\right\}\left\{\prod_{p=2}^{2m-2}(n-p)\right\}\psi^{m-1}\right]\nonumber\\
	&\ &\hspace{1cm}+\sum_{m,j=2}^{k}\Biggl[a_ma_j(n-1)\left\{\prod_{p=2}^{2m-2}(n-p)\right\}\left\{\prod_{p=1}^{2j-2}(n-p)\right\}\nonumber\\
	&\ &\hspace{4cm}\times \frac{j(n-1)-(m-1)(n+1)}{j}\psi^{j+m-2}\Biggr]\ . \label{eq:K}
\end{eqnarray}
The factor other than $K[\psi]$ in (\ref{eq:hpsi}) are manifestly positive, 
so the sign of $K[\psi]$ determines that of $T^{'}$. 
However, from (\ref{eq:K}), it is clear that sign of $K[\psi]$ depends on dimensions and Lovelock coefficients $a_m$. 
Therefore, if Lovelock black holes have the ghost instability
 depends on dimensions and Lovelock coefficients.

However, as we will show later, $T^{''}$ has a negative region in even-dimensions 
if Lovelock black holes are sufficiently small. 
Therefore, even if $T^{'}$ are always positive and consequently
 Lovelock black holes have no ghost instability, they have the dynamical instability 
as long as the ADM mass is sufficiently small.
 
Substituting the explicit form of $W[\psi]$ into the formula (\ref{T_double_prime}),
 we get
\begin{eqnarray}
	T^{''}=r^{n-3}\frac{L[\psi]}{\left(1+\sum_{m=2}^{k}\left[a_m\left\{\prod_{p=1}^{2m-2}(n-p)\right\}\psi^{m}\right]\right)^3}\ . \label{eq:dhpsi}
\end{eqnarray}
Here, the lowest and the leading term of $L[\psi]$ is 
\begin{eqnarray}
	L[\psi]&=&(n-1)(n-2)+\cdots\nonumber\\
	&\ &\ +\frac{a_k^4}{k^2}\left\{\prod_{p=1}^{2k-2}(n-p)^4\right\}(n-(2k-1))(n-(3k-1))\psi^{4k-4} \ . \label{eq:L}
\end{eqnarray}
We note that the highest order $k=[(D-1)/2]$ is related to
 dimensions as $n=2k-1$ in odd-dimensions and $n= 2k$ in even-dimensions.
In odd-dimensions, the leading term disappears.
Hence, we cannot say anything in general. Hence, we consider only
even-dimensions.

Let us examine the sign of $L[\psi]\ (\psi\geq0)$. 
If $n=2k$, the coefficient of the lowest term is positive and 
that of the leading one of (\ref{eq:L}) is negative. 
Therefore, $L[\psi]>0$ near $\psi=0$ and $L[\psi]<0$ for large $\psi$. This means that there exists 
roots of $L[\psi]=0$ because $L[\psi]$ is a continuous function. 
Let $\psi_0$ be the lowest positive root. 
If $\psi_H<\psi_0$, then $L[\psi]>0$ for $0\leq\psi\leq \psi_H$,
 and hence we conclude $T^{''}>0$ for $r>r_H$. 
While, if $\psi_H>\psi_0$, then there exists a region $L[\psi]<0$ in 
the range $\psi_0\leq\psi\leq \psi_H$.
 Thus, there exists a region $T^{''}<0$ outside the horizon $r>r_H$. 
Therefore, black holes are stable 
if $\psi_H < \psi_0$ and unstable if $\psi_H>\psi_0$. 
Since $\psi_H$ becomes larger as $\mu$ becomes smaller,
we conclude that there exist a critical mass below which black holes become 
unstable.

To conclude this section,
by considering tensor perturbations, we can say that 
``When the ADM mass is sufficiently small,
Lovelock black holes in even-dimensions have the ghost instability or
the dynamical instability; that is, small Lovelock black holes are unstable
in even-dimensions ".

%===============================================================%
%************************ SECTION IV ***************************%
%===============================================================%
\section{Stability analysis for vector perturbations}
\label{seq:4}

In this section, we consider the stability of Lovelock black holes under
vector perturbations. 

Master equation for vector perturbation is given by~\cite{Takahashi}; 
\begin{eqnarray}
	-\partial_{r^{*}}^2\Psi+V_v(r)\Psi=\omega^2\Psi \label{}
\end{eqnarray}     
where 
\begin{eqnarray}
V_v(r)=\frac{1}{\frac{1}{r\sqrt{T^{'}}}}f\partial_r \left(f\partial_r\frac{1}{r\sqrt{T^{'}}}\right)+\left(\frac{\gamma_v}{n-1}-\kappa\right)\frac{fT^{'}}{rT}\ .
\label{}
\end{eqnarray}
Here, $\Psi$ is the master variable and $\gamma_v$ are eigenvalues of vector harmonics 
with $\gamma_v=\ell(\ell+n-1)-1$ $(\ell\geq1)$ for $\kappa=1$ and 
 non-negative real numbers for $\kappa=0,-1$.  
This equation is obtained provided that tensor perturbations have no ghost instability. 
Moreover, $T(r)$ is always positive, because
we assumed the conditions (\ref{eq:conditions}), the positivity of $\mu$ ,
 and the asymptotically flat spherical symmetric branch.   

In this section, we show that Lovelock black holes are stable under vector perturbations
 as long as they have no ghost instability under tensor perturbations. 
In order to prove this statement, we again use the S-deformation approach. 
We define ${\cal H}=-d^2/dr^{*2}+V_v(r)$ and the inner product (\ref{inner_product}). 
Then, as in the last section, we can find smooth function $S$ such that 
\begin{eqnarray}
	(\varphi,{\cal H}\varphi)=\int_{I} (|D\varphi|^2+\tilde{V}|\varphi|^2)dr^{*},\nonumber
\end{eqnarray}
for any $\varphi$ . Here, we have defined
\begin{eqnarray}
	D=\frac{d}{dr^{*}}+S  \ , \quad 
	\tilde{V}=V_v+f\frac{dS}{dr}-S^2  \ .\nonumber
\end{eqnarray}
Following the paper \cite{Gleiser:2005ra}, we choose $S$ to be
\begin{eqnarray}
S=-f\frac{d}{dr}\ln(\frac{1}{r\sqrt{T^{'}}})\ . 
\label{}
\end{eqnarray}
Then, ${\tilde V}$ can be calculated as 
\begin{eqnarray}
\tilde{V}=\frac{\gamma_v-(n-1)\kappa}{n-1}\frac{fT^{'}}{rT} \ .
\label{}
\end{eqnarray}
In this new potential ${\tilde V}$, apparently
$\gamma_v-(n-1)\kappa>0$, $f>0$ and $T>0$.  We also assumed $T^{'}>0$, so 
it is clear that ${\tilde V}$ is always positive. Therefore, the inner product $(\varphi,{\cal H}\varphi)$ is positive for any $\varphi$. 
In particular, it is true for the lowest energy state, 
hence the lowest energy is positive. 
This implies that black holes are stable under vector perturbations. 

To summarize this section, we can say that 
`` if there is no ghost instability in tensor perturbation, 
Lovelock black holes are stable under vector  perturbations". 

%===============================================================%
%************************ SECTION V ***************************%
%===============================================================%
\section{Stability analysis for scalar perturbations}
\label{seq:5}

In this section, we examine the stability of Lovelock black holes
under scalar perturbations. 

In the previous paper, we have derived the master equation for 
scalar perturbations~\cite{Takahashi}. 
  Using the master variable $\Psi$, we can write down the master equation
\begin{eqnarray}
-\partial_{r^*}^2\Psi+V_s(r)\Psi=\omega^2\Psi    \ .
\label{}
\end{eqnarray}
Here, the effective potential reads
\begin{eqnarray}
V_s(r)&=&2\gamma_sf\frac{(rNT)^{'}}{nNTr^2}\nonumber\\
&\ &\hspace{0.5cm}-\frac{f}{N}\partial_r\left(f\partial_rN \right)+2f^2\frac{N^{'2}}{N^{2}}-\frac{f}{T}\partial_r(f\partial_rT)+2f^2\frac{T^{'2}}{T^2}+2f^2\frac{N^{'}T^{'}}{NT}\ ,
\label{S_potential}
\end{eqnarray}
where we have defined 
\begin{eqnarray}
N(r)=\frac{-2nf+2\gamma_s+nrf^{'}}{r\sqrt{T^{'}}} \ .
\label{}
\end{eqnarray}
For scalar perturbations, eigenvalues of scalar harmonics $\gamma_s$  
 are given by $ \gamma_s=\ell(\ell+n-1)$ for 
$\kappa=1$ and positive real numbers for $\kappa=0,-1$.
The above master equation is obtained by assuming $T'>0$.
 Hence, tensor perturbations have no ghost instability. 

Note that we will consider an asymptotically flat spherically symmetric branch 
   with positive mass as in section \ref{seq:2}.  
   In this branch, $T(r)$ is always positive. 
  
\subsection{Criterion for Instability}
  
  In this subsection,  we show that black holes are 
  unstable if $2T^{'2}-TT^{''}$ has a negative region outside  the horizon. 

In order to prove this statement, we can use the S-deformation approach. 
Here, we choose $S$ as 
\begin{eqnarray}
S=f\partial_r(\ln N)+f\partial_r(\ln T)\ ,
\label{}
\end{eqnarray}
then the second line of (\ref{S_potential}) canceled and ${\tilde V}$ becomes 
\begin{eqnarray}
{\tilde V}&=&2\gamma_sf\frac{(rNT)^{'}}{nNTr^2}\nonumber\\
&=&\frac{2\gamma_sf}{nr}\left[\frac{2(\gamma_s-n\kappa)}{2(\gamma_s-n\kappa)+\frac{n(n+1)\mu}{T}}\frac{T^{'}}{T}-\frac{1}{2}\frac{T^{''}}{T^{'}}\right]\nonumber\\
&<&\frac{\gamma_sf}{nrTT^{'}}\left[2T^{'2}-TT^{''}\right]\ .
\label{}
\end{eqnarray}
Note that we used the assumption $T>0$, $T^{'}>0$ and $\mu >0$ in the last inequality. 

Now, let us prove the statement ``if $2T^{'2}-TT^{''}$ has a negative region, 
black holes are unstable". 
In order to do that, we use the inequality 
\begin{eqnarray}
\omega^2_0\leq \frac{(\varphi,{\cal H}\varphi)}{(\varphi,\varphi)}\ , 
\label{}
\end{eqnarray}
where $\omega^2_0$ is the lowest eigenvalue. 
This inequality is true for arbitrary test function $\varphi$, 
so we choose $\varphi$ as the smooth function that has compact support in the region 
where $2T^{'2}-TT^{''}$ is negative. Then, $\omega^2_0$ can be bounded as
\begin{eqnarray}
\omega^2_0&\leq&(\varphi,{\cal H}\varphi)/(\varphi,\varphi)\nonumber\\
       &=&\frac{1}{(\varphi,\varphi)}\int dr^{*}\left[\left|D\varphi \right|^2+{\tilde V}\left|\varphi\right|^2 \right]\nonumber\\
       &<&\frac{1}{(\varphi,\varphi)}\left[\int dr^{*}\left|D\varphi \right|^2+\gamma_s \int dr^*\frac{f}{nrTT^{'}}\left(2T^{'2}-TT^{''}\right)\left|\varphi\right|^2 \right]\ .
\label{E_0_bound}
\end{eqnarray}
We assume $T>0$, $T^{'}>0$ and choose $\varphi$ as the smooth function
 which has compact support in the region $2T^{'2}-TT^{''}<0$, 
so the first term in (\ref{E_0_bound}) must be positive and the second term
in (\ref{E_0_bound}) must be negative. 
Therefore, by taking the limit $\gamma_s=\ell(\ell+n-1)\rightarrow \infty$, 
the last line of Eq.(\ref{E_0_bound}) becomes negative, which means 
the lowest eigenvalue $\omega_0^2$ is negative. Hence, black holes are dynamically unstable. 

In summary, we can say that 
``If $T^{'}$ has a negative region, black holes have the ghost 
instability under tensor perturbations. Even if $T^{'}$ is always positive, 
Lovelock black holes have the dynamical instability under scalar perturbations
 if $2T^{'2}-TT^{''}$ has a negative region". 
Note that we can not say that black holes are stable
 even if $2T^{'2}-TT^{''}$ is always positive.

\subsection{Instability of Small Lovelock Black Holes in Odd-dimensions}

Now let us check the sign of $T^{'}$ and $2T^{'2}-TT^{''}$. 
We have already shown that Lovelock black holes are unstable 
in even-dimensions under tensor perturbations, so we concentrate on odd-dimensions. 
In order to examine the sign of these functions, it is convenient to express
 these functions by $\psi$. 
The formula (\ref{T_prime}) reads
\begin{eqnarray}
	T^{'}(r)=r^{n-2} \frac{K[\psi]}{\partial_{\psi}W[\psi]}\ , 
\end{eqnarray}
where 
\begin{eqnarray}
K[\psi]=(n-1)(\partial_{\psi}W)^2-(n+1)W\partial_{\psi}^2W\ .
\label{}
\end{eqnarray}
Similarly, the formula (\ref{T_double_prime}) can be written as
\begin{eqnarray}
2T^{'2}-TT^{''}=\frac{r^{2n-4}}{(\partial_{\psi}W)^2}M[\psi]\ ,\label{2T_prime_square}
\end{eqnarray}
where
\begin{eqnarray}
M[\psi]&=&n(n-1)(\partial_{\psi}W)^4-3n(n+1)W\partial_{\psi}^2W(\partial_{\psi}W)^2\nonumber\\
&\ &\hspace{2cm}+3(n+1)^2W^2(\partial_{\psi}^2W)^2-(n+1)^2W^2\partial_{\psi}W\partial_{\psi}^3W
\label{}\ .
\end{eqnarray}
Apparently, the signs of $T^{'}$ and $2T^{'2}-TT^{''}$ are determined 
by $K[\psi]$ and $M[\psi]$. So we have to  
 check the signs of $K[\psi]$ and $M[\psi]$. 

First, we consider the sign of $T^{'}$ which has been already expressed by $\psi$
 in (\ref{eq:hpsi}) and (\ref{eq:K}). 
We here consider odd-dimensions $n=2k-1$, then we have  
\begin{eqnarray}
K[\psi]=2(k-1)+\cdots+2\frac{\alpha_{k-1}^2(k-2)-2\alpha_{k}\alpha_{k-2}(k-1)}{(k-1)(k-2)}\psi^{2k-4}\ .
\label{scalar_K}
\end{eqnarray}
where $\alpha_m=a_m\left\{\prod_{p=1}^{2m-2}(n-p)\right\}$. 
Furthermore, for $n=2k-1$, $M[\psi]$ can be expressed by $\psi$ as 
\begin{eqnarray}
M[\psi]=2(2k-1)(k-1)+\cdots -6a_{k}^2\frac{\alpha_{k-1}^2(k-2)-2\alpha_{k}\alpha_{k-2}(k-1)}{(k-1)(k-2)}\psi^{4k-6} \ .
\label{scalar_M}
\end{eqnarray}
The most important point is that 
the signs of the coefficients of the leading term in (\ref{scalar_K})
 and (\ref{scalar_M}) are opposite. 

First, we assume $\alpha_{k-1}^2(k-2)-2\alpha_{k}\alpha_{k-2}(k-1)$ is negative. In this case, $K[\psi]$ is positive near $\psi=0$ and negative for large $\psi$, 
so $K[\psi]$ has positive roots. Let the lowest root be $\psi_0$. 
Note that $\psi$ moves in the range $(0,\psi_H]$. Then, if $\psi_H>\psi_0$, 
$K[\psi]$ has a negative region and $T^{'}$ does so, which means Lovelock black holes have the ghost instability. Note that $\psi_H$ becomes larger as $\mu$ becomes smaller. 
Then, if $a_{k-1}^2(k-2)-2a_{k}a_{k-2}(k-1)$ is negative, small Lovelock black holes have the ghost instability. 

Conversely, we assume  $\alpha_{k-1}^2(k-2)-2\alpha_{k}\alpha_{k-2}(k-1)$ is positive.
 In this case, the coefficient of the leading term of $K[\psi]$ is positive, so 
whether $K[\psi]$ has positive root or not is obscure. If this has a positive root, we can say that small Lovelock black holes have the ghost instability by using the same logic as the last paragraph.  However, even if $K[\psi]$ has no positive root, $M[\psi]$ must have positive roots; this is because $M[\psi]$ becomes positive near $\psi=0$ and must be negative for large $\psi$. Let the lowest positive root be $\psi_1$. Using this definition,we can say  Lovelock black holes have the dynamical instability if $\psi_H>\psi_1$, 
namely, if Lovelock black holes are sufficiently small. 

In any case, assuming black holes are sufficiently small, Lovelock black hole have 
the ghost instability or the dynamical instability in odd-dimensions.

To conclude this section, we can say that 
``small Lovelock black holes are unstable in odd-dimensions". 

%===============================================================%
%************************ SECTION VI ***************************%
%===============================================================%
\section{Example: Einstein-Gauss-Bonnet Theory}
\label{seq:6}

In this section, to illustrate our general statement, 
we take Einstein-Gauss-Bonnet theory as an example. 
The Einstein-Gauss-Bonnet theory corresponds to $k=2$ and
generic for $n=3$ and $n=4$.

The action for Einstein-Gauss-Bonnet theory is given by
\begin{eqnarray}
   L = R + \frac{a_2}{2} \left( R^2-4R_{\mu}{}^{\nu}R_{\nu}{}^{\mu}
   +R_{\mu \nu}{}^{\lambda \rho}R_{\lambda \rho}{}^{\mu \nu} \right) \ .
\end{eqnarray}
The stability of this model has been already analyzed in \cite{Dotti:2004sh}
and \cite{Gleiser:2005ra}. However, the derivation presented here is more
explicit and hence transparent.

Substituting the concrete form of $W[\psi ]$ into $T^{'}$, 
we can get 
\begin{eqnarray}
	K[\psi]=
      (n-1)+2a_2(n-1)(n-2)(n-3)\psi+a_2^2(n-1)^2(n-2)^2(n-3)\psi^2
      \ . \label{eq:k2_hpsi}
\end{eqnarray}
As we can see from (\ref{eq:k2_hpsi}), $T^{'}$ is always positive 
both for $n=3$ and $n=4$ because we are considering cases $\psi>0$ and $a_2 >0$. 
Similarly, substituting the explicit form of $W[\psi]$ into $T^{''}$, we obtain 
\begin{eqnarray}
L[\psi]&=&
      (n-1)^4(n-2)^4(n-3)(n-5)a_2^4\psi^4+4(n-1)^3(n-2)^3(n-3)(n-5)a_2^3\psi^3
      \nonumber\\
	&\ &\hspace{1cm}+2(n-1)^2(n-2)^2(5n^2-25n+42)a_2^2\psi^2
      \nonumber\\
	&\ &\hspace{1.5cm}+12(n-1)(n-2)(n^2-3n+4)a_2\psi+4(n-1)(n-2) \ . 
      \label{criterion:L}
\end{eqnarray}
Finally,  $M[\psi]$ which determines the sign of 
$2T^{'2}-TT^{''}$ can be calculated as  
\begin{eqnarray}
M[\psi]&=&\frac{1}{4}(n-1)^3(n-2)^4(n-3)a_2^4\psi^{4}
+(n-1)^4(n-2)^3(n-3)a_2^3\psi^{3} \nonumber\\
&\ & +\frac{3}{2}(n-1)^2(n-2)^2(n^2-5n+2)a_2^2\psi^2
\nonumber\\
&\ &\hspace{1cm} + n(n-1)(n-2)(n-7)a_2\psi
+n(n-1)\ .
\label{k2_M}
\end{eqnarray}
The functions $L[\psi]$ and $M[\psi]$ give criterions for the instability of
tensor and scalar perturbations, respectively.

Let us first consider 6-dimensional Einstein-Gauss-Bonnet black holes
 corresponding to $n=4$.
Then, substituting  $n=4$ into (\ref{k2_M}), we obtain
\begin{eqnarray}
M[\psi]=12(-1+3a_2\psi+9a_2^2\psi^2)^2>0\ .
\label{}
\end{eqnarray}
Therefore, black holes are stable under scalar perturbations. 
Next, we need to check the stability of Lovelock black holes under tensor perturbations.
In the case of $n=4$, the formula (\ref{criterion:L}) becomes
\begin{eqnarray}
 L[\psi] = -1296 a_2^4 \psi^4 - 864 a_2^3 \psi^3 + 1584 a_2^2 \psi^2 + 576 a_2 \psi +24 \ .
\end{eqnarray}
Clearly, the coefficient of the leading term of $L[\psi]$ is negative.
 Thus, for large $\psi$, $L[\psi]$ is negative. 
 And, $L[\psi]$ is positive near $\psi\sim 0$ because $L[0]=24>0$. 
 Hence, there must be a root somewhere between. 
 Indeed, $L[\psi]$ becomes zero at
 \begin{eqnarray}
\psi_0=\frac{1}{6a_2}(-1+\sqrt{15}+\sqrt{10})\ .
\label{}
\end{eqnarray}
 Therefore, $L[\psi]$ is always positive in the range $0<\psi<\psi_0$ 
 and always negative in the range $\psi>\psi_0$.
As we explained, $\psi$ moves in the range $0<\psi\leq\psi_H$. 
Therefore, if $\psi_H>\psi_0$, there exists the region $T^{''}<0$, 
which means Lovelock black holes are dynamically unstable. 
Furthermore, from (\ref{eq:rh}), the inequality $\psi_H>\psi_0$ yields 
\begin{eqnarray}
	\mu<\frac{3\sqrt{6}\left(1+\sqrt{15}+\sqrt{10} \right)}
      {\left(-1+\sqrt{15}+\sqrt{10} \right)^{3/2}}a_2^{3/2} \equiv \mu_{c}\ .
\end{eqnarray}
This proves that Lovelock black holes with the mass less than $\mu_c$
 are unstable in 6 dimensions. 

Now, let us consider 5-dimensional Einstein-Gauss-Bonnet black holes
corresponding to $n=3$.
In the case of $n=3$, the formula (\ref{criterion:L}) reads
\begin{eqnarray}
L[\psi] = 96 a_2 \psi^2 +96 a_2 \psi +8 \ .
\end{eqnarray}
Since all the coefficient of $L[\psi]$ are positive and 
we are considering positive $\psi$,  $L[\psi]$ is always positive 
and hence $T^{''}$ is always positive. 
Thus, Lovelock black holes in 5-dimensions are stable under tensor perturbations.
 However, for $n=3$, $M[\psi]$ becomes 
\begin{eqnarray}
M[\psi]=6(1-4a_2\psi-4a_2^2\psi^2)\ .
\label{}
\end{eqnarray}
From this equation, it is easy to see that $M[\psi]=0$ has a positive solution;
 that is $\psi_0=\frac{\sqrt{2}-1}{2a_2}$. 
Then, $M[\psi]$ is positive in the range $0<\psi<\psi_0$ and 
negative in the range $\psi>\psi_0$. 
Since $\psi$ moves in the range $0<\psi\leq\psi_H$, 
$2T^{'2}-TT^{''}$ has a negative region if $\psi_H>\psi_0$.  
Using the solution for $\psi_H$ (\ref{eq:rh1}), 
we can rewrite the inequality $\psi_H>\psi_0$ as
\begin{eqnarray}
a_2<\mu<(\sqrt{2}+1)^2a_2\ .
\label{}
\end{eqnarray}
Note that the lower bound came from the condition for the existence of the horizon as
 we have explained in Sec.\ref{seq:2}. 
Hence, 5-dimensional Lovelock black holes with the mass  
in the above range are dynamically unstable under scalar perturbations.

%===============================================================%
%************************ SECTION VII ***************************%
%===============================================================%
\section{Conclusion}
\label{seq:7}

We have studied the stability of static black holes in Lovelock theory
which is a natural higher dimensional generalization of Einstein theory.   
We have shown that there exists the instability of Lovelock black holes 
with small mass under tensor perturbations in even-dimensions and
 under scalar perturbations in odd-dimensions. 
Lovelock black holes are stable under vector perturbations 
as long as they do not have ghost instability under tensor perturbations. 
Remarkably, the instability is stronger on short distance scales,
which is different from the usual instability for which the onset of the
instability becomes a bifurcation point.
Hence, the instability we have discussed in this paper is catastrophic
in the sense that there is no smooth descendant.  
Curiously, the similar instability also appears
in the Gauss-Bonnet cosmology~\cite{Kawai:1998ab}. 
In spite of this unusual nature of the instability, 
it is interesting to investigate the fate of the catastrophic instability.
This issue is very important because black holes lose their mass due to
the Hawking radiation and eventually become unstable. 

It is worth examining the more profound meaning of 
this instability; that is, why Lovelock black holes are unstable. 
Especially, it is very interesting to find the reason why 
black holes have the instability under tensor perturbations in even-dimensions and 
under scalar perturbations in odd-dimensions. 

Related to the above, it is intriguing to find 
the thermodynamical meaning of the universal function $T(r)$. 
As was shown in this paper, this function governs the dynamical stability of black holes. 
Therefore, if $T(r)$ has thermodynamical meaning, the relation between 
thermodynamical~\cite{Cai:2001dz, Dehghani:2005vh} and dynamical instability might be revealed. 

It is interesting to investigate if the instability we found also
exists for asymptotically AdS cases from the point of view of the AdS/CFT 
 correspondence~\cite{Ge:2009ac,Shu:2009ax,deBoer:2009gx},
 in particular, in relation to stability of holographic 
 superconductors~\cite{Gregory:2009fj,Kanno:2010pq}. 

\begin{acknowledgements}
JS is supported by  the
Grant-in-Aid for  Scientific Research Fund of the Ministry of 
Education, Science and Culture of Japan No.22540274, the Grant-in-Aid
for Scientific Research (A) (No. 22244030), the
Grant-in-Aid for  Scientific Research on Innovative Area No.21111006
and the Grant-in-Aid for the Global COE Program 
``The Next Generation of Physics, Spun from Universality and Emergence".
\end{acknowledgements}

\end{document}